\documentclass[manuscript]{acmart}
\usepackage{tabularx}
\usepackage{hyperref}
\usepackage{todonotes}
\usepackage{balance}
\usepackage{booktabs}
\usepackage{natbib}
\usepackage{caption}
\usepackage{subcaption}
\AtBeginDocument{%
  \providecommand\BibTeX{{%
    \normalfont B\kern-0.5em{\scshape i\kern-0.25em b}\kern-0.8em\TeX}}}

\setcopyright{acmcopyright}
\copyrightyear{2022}
\acmYear{2022}
\acmDOI{XXXXXXX.XXXXXXX}

\acmConference[A Workshop at CHI'22 ]{CHI'22 Workshop: The Future of Emotion in Human-Computer Interaction}{April 13th--April 14th, 2022}{New Orleans, LA, USA}
\acmBooktitle{CHI'22 Workshop: The Future of Emotion in Human-Computer Interaction, April 13th--April 14th, 2022, New Orleans, LA}
\acmPrice{XX.XX}
\acmISBN{978-1-4503-XXXX-X/18/06}
\begin{document}

%\title{Emotion Reaction: How to Deal With Users' Negative Moods?}
\title{How Should Voice Assistants Deal With Users' Emotions?}

\author{Yong Ma}
\affiliation{%
\institution{LMU Munich}
\city{Munich}
\country{Germany}}
\email{yong.ma@ifi.lmu.de}

 \author{Heiko Drewes}
\affiliation{%
\institution{LMU Munich}
\city{Munich}
\country{Germany}}
\email{heiko.drewes@ifi.lmu.de}

\author{Andreas Butz}
\affiliation{%
\institution{LMU Munich}
\city{Munich}
\country{Germany}}
\email{andreas.butz@ifi.lmu.de}

\renewcommand{\shortauthors}{Submission to CHI'2022 Workshop: please do not circulate!}

\begin{abstract}
%With the advancement of speech technologies, Voice Assistants (VAs) play a increasingly significant role in our daily lives. They can build the communication bridge between users and computers. Generally, VAs exist in both mobile and stationary devices and users can easily interact with them through voice or speech command. VAs can also perceive users' emotions by semantic analysis or speech emotion recognition. However, it becomes quite challenging to respond to users' emotion, especially users' negative emotion. Based on the emotion reaction strategies from humans, we switched the roles of the users and the VIs. How to respond to the emotional stimulus becomes our main research issues. We designed three avatar emojis (angry, sad and frightened) that can express the specific emotions through animation and voice. 52 participants were invited to conduct our user study and they tried to make the emojis become a desired emotional state (positive emotions) through emotional voice input. The results show that users predominantly used the neutrally emotion to react these three negative emotions and it exists difference in genders emotion reaction. Yet, differences in response to negative emotions by genders are unclear, since our experiments only used male voices as emotional stimuli. 
There is a growing body of research in HCI on detecting the users' emotions. Once it is possible to detect users' emotions reliably, the next question is how an emotion-aware interface should react to the detected emotion. In a first step, we tried to find out how humans deal with negative emotions of an avatar. The hope behind this approach was to identify human strategies, which we can then mimic in an emotion-aware voice assistant.
We present a user study in which participants were confronted with an angry, sad, or frightened avatar. Their task was to make the avatar happy by talking to it. We recorded the voice signal and analyzed it. The results show that users predominantly reacted with neutral emotion. However, we also found gender differences, which opens a range of questions.
\end{abstract}

\begin{CCSXML}
<ccs2012>
<concept>
<concept_id>10003120.10003121.10003124.10010870</concept_id>
<concept_desc>Human-centered computing~Natural language interfaces</concept_desc>
<concept_significance>500</concept_significance>
</concept>
</ccs2012>
\end{CCSXML}

\ccsdesc[500]{Human-centered computing~Natural language interfaces}

\keywords{Emotion Reaction, Negative Emotions, Avatar Emojis}

\maketitle

\section{Introduction}

Voice Assistants (VAs), which are embedded in smartphones (e.g., Siri) or smart home devices (e.g., Alexa), have a growing number of users. The advantage of VAs, compared to classical interaction with keyboard, mouse and display, is that they do not demand visual attention and leave the hands free for other tasks~\cite{sayago2019voice}. Despite the 'artificial intelligence' in VAs, users consider current voice assistants stupid, as they are not fully-fledged conversational partners. One reason for this is a missing awareness of the users' emotions.

With the advancements in neural networks and machine learning it is possible to recognize users' emotions in speech~\cite{akccay2020speech,fahad2021survey}. However, even if it is possible to reliably detect emotion, it is not clear how a VA should react. This is the research question of our study.

Some emotions, such as sadness, anger, and fear, are classified as negative, and an often-heard idea is to develop VAs, which turn such emotion into a positive one, i.e., make the user happy. The question is how a VA should behave and how the VAs voice should sound in an emotional response to achieve this goal. %VAs need strategies and rules to deal with users' emotions.
To get closer to an answer, we decided to study how humans would try to achieve this. For a study we set up a website with animated emojis talking in negative emotions. We asked the participants to cheer the emojis up by talking, and we recorded their voices. We analyzed the recorded voice samples with Vokaturi\footnote{https://vokaturi.com/}, an emotion detector based on neural networks, and also with classical methods such as RMS and ZCR. The predominant result from Vokaturi was 'neutral' and there were not many differences in the voices for different moods of the emoji. However, we found surprising differences in strategy by the participants' gender. 
Our study did therefore not deliver conclusive answers, but created questions for further research, especially on the gender issue.

%Voice Assistants (VAs), which are embeded in our smartphone or smart home devices, have sparked the widespread use in our daily lives. VAs exist in both mobile and stationary devices and users can be allowed to easily interact with them through voice or speech command. More specifically, users can interact with VAs by voice or speech and VAs can recognize the users’ spoken commands after users’ saying. Then VAs will give their answers based on their understanding and yet they won’t say anything afterwards like humans’ communication. Generally, current VAs cannot perceive users’ emotion via voice and express their own emotions when they talk with users. 
%With the advancement of speech signal processing and speech emotion recognition, it is possible to recognize users' emotion by speech input\cite{akccay2020speech,fahad2021survey}. Yet, it becomes critically challenging to deal with users emotions. 

\section{Related Work}

There is a growing body of research on how to detect emotion in speech. One possibility to detect emotion is speech emotion recognition (SER), which analyzes \emph{how} something was said. A good overview on this approach is provided by \citet{schuller2018speech}. Another possibility is the semantic analysis of natural language~\cite{maulud2021state}, which analyzes \emph{what} was said.

Research on how to react to emotion is mainly done by psychologists, but it typically investigates the communication between humans.
%Psychologists proposed many methods which can be used to react to the others' emotions.
\citet{li2021emoelicitor} worked on emotional reaction by focusing on the feelings of the user and presented the 'EmoElicitor' model to elicit the particular emotions of users. 
\citet{clos2017predicting} tried to predict the emotional reaction of readers of social network  posts. 
Other researchers concluded that their approach outperforms other approaches they took as a baseline, e.g., estimating emotion from text by using 'EmoLex'~\cite{mohammad2010emotions}. 
\citet{thornton2017mental} studied whether humans can predict the future emotion of others from the currently observed emotion through a learned mental model. %Yet, it is still hard for VAS to react to users' emotions based on these methods. Is it possible to do research in another side? Changing the roles of the VAs and users gives us another option for conducting our user research.

%It is crucially important for VAs to perceive and appropriately react to users' emotions. With the advancement of VAs, it is possible to detect users' emotions. How to response to users' emotion becomes the challenging issues for the current VAs.

%\subsection{Voice Assistant}
%Compared to the traditional interfaces such as such as keyboard, mouse, and touchscreen, VAs keep the users' hands and eyes free for other tasks\cite{sayago2019voice} when they used the VAs for interaction. 
%With the development of speech emotion recognition\cite{schuller2018speech}, the sematic analysis of natural language processing\cite{maulud2021state}, it becomes feasible to sense users' emotions via context and voice. However, it is extremely difficult for VAs to select the appropriate emotions to respond to users. 

%\subsection{Emotion Reaction}

In the field of HCI, some researchers proposed self-reported and concurrent expression which can help computers to efficiently sense, recognize and respond to human communication of emotions~\cite{picard2000toward,picard2002computers,picard2003computers}. \citet{pascual2007emotional} used a sequential model of emotional processing and an accompanying measure to conduct an emotional transformation study involving 310 clinical and 130 sub-clinical cases~\cite{pascual2018clients}. 
%Nevertheless, it still exists certain issues to deal with users emotions in VAs. Concretely, it is still incredibly difficult for VAs to respond to users' emotions without any prior knowledge, even while VAs can understand users' emotions. It means VAs need to take certain strategies or follow certain rules to deal with users' emotions. 

\section{Study}
\begin{figure*}[bh]
%\begin{minipage}[t]{0.45\textwidth}
\centering
  \includegraphics[width=0.33\textwidth]{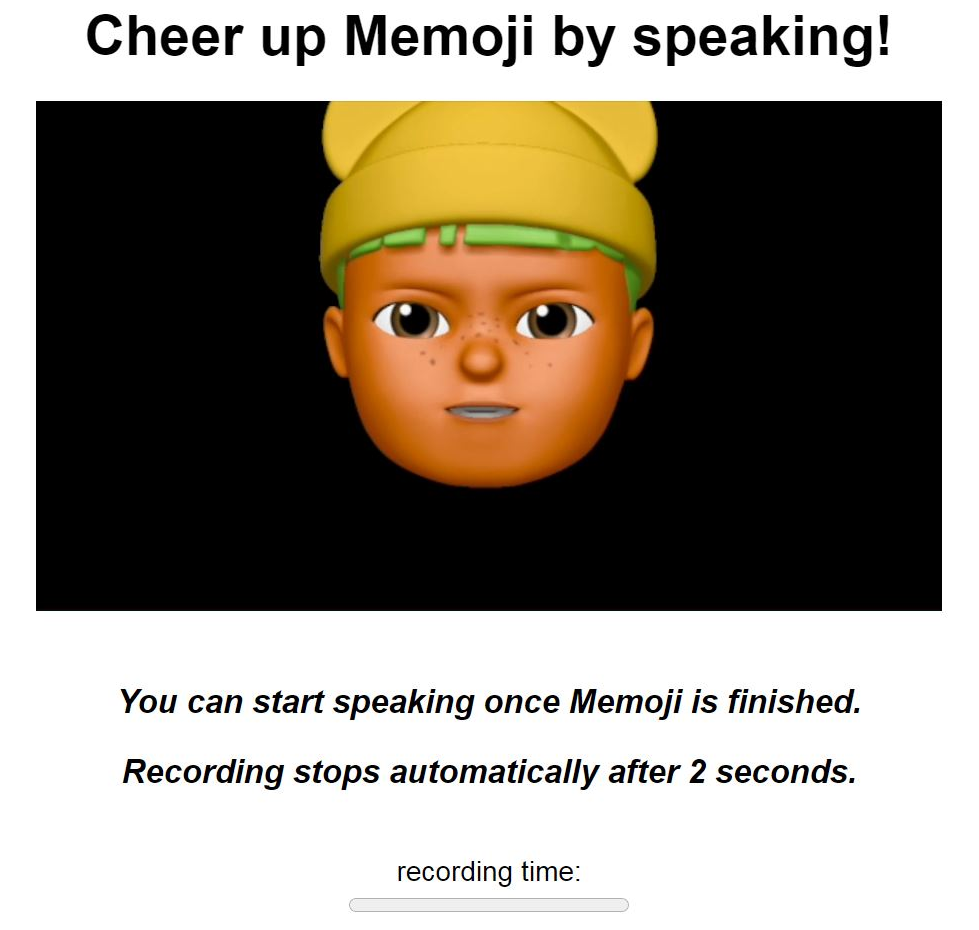}\hfill
  \includegraphics[width=0.33\textwidth]{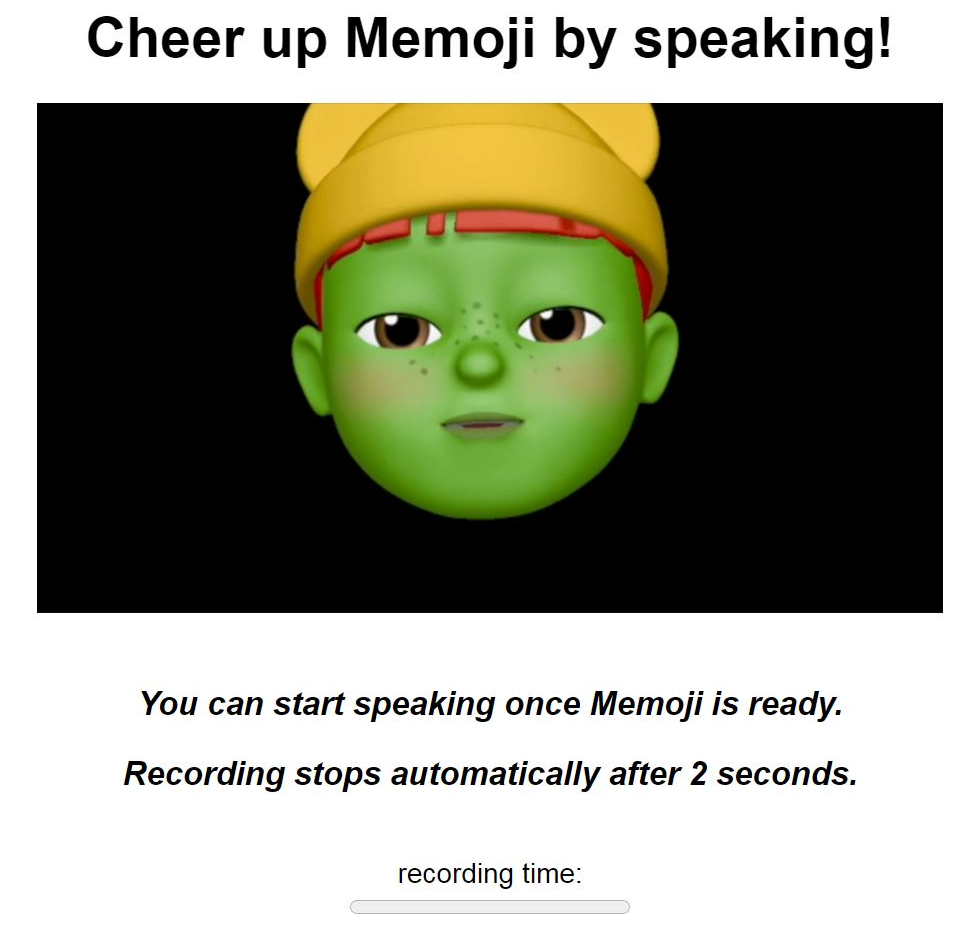}\hfill
  \includegraphics[width=0.33\textwidth]{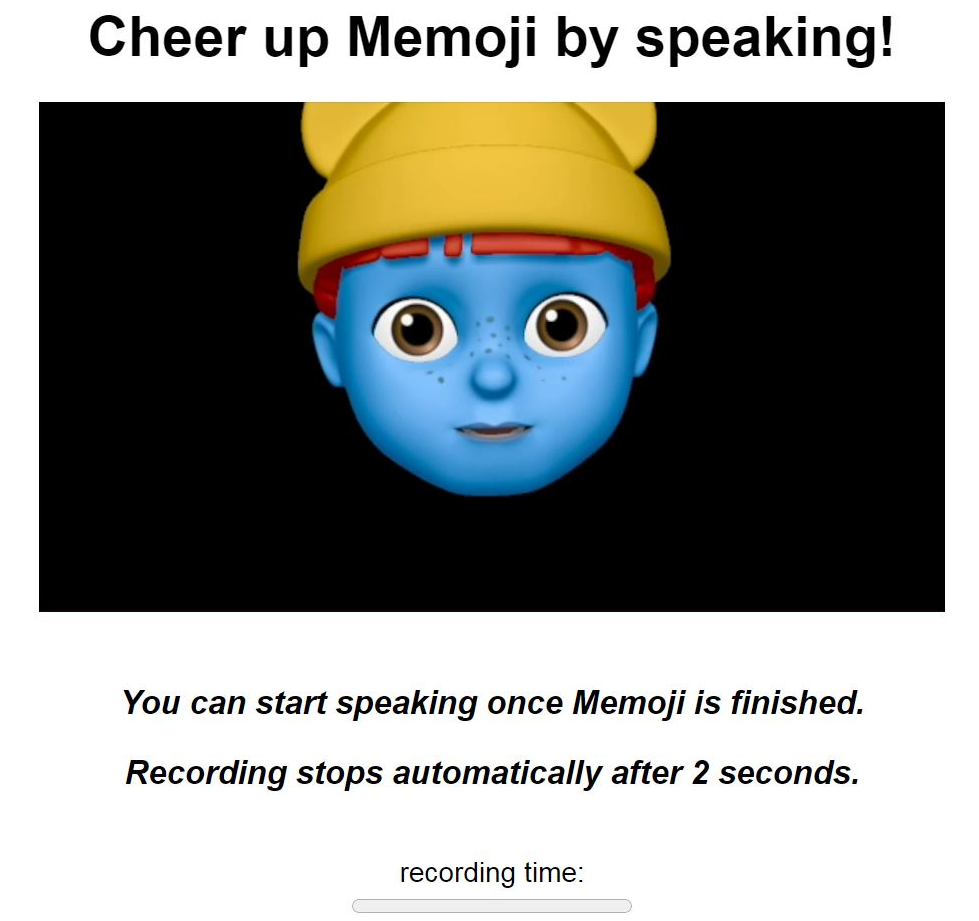}
  \caption{The web page offered emojis in three moods - angry, sad, fearful. The emoji talked to the participants and waited for an answer, which we recorded. This was repeated with different emoji utterances five times for each emotion.}\label{fig:webpage}
%  \vspace{-1em}
%  \end{minipage}
\end{figure*}
To understand how VAs should react to users' emotion, we used the approach of role-swapping. In other words, the roles of VAs and individuals are switched in our study. We designed a website  and invited participants with the web link. Besides a privacy statement, instructions, and a microphone test, the website presented animated avatars (see \autoref{fig:webpage}) in three different emotional states -- sad, angry, and fearful. The participants' task was to change the avatars' emotion into happiness by speaking to them. For every emotion, the conversation consisted of five avatar utterances and corresponding user answers. After each answer the avatar become gradually happier, independently of the user's actual answer. We recorded the first two seconds of every answer.
After the recordings of each emotion we asked 'How sad was the emoji's voice?" and "How sad was the emoji's face?" and the analog for the other emotions to validate the emotional input. We also asked for suggestions how to cheer up people in this mood to collect users' intentions.

\section{Results and Analysis}
The study was conducted during three weeks and involved 52 participants. Six participants did not complete the study and therefore, 46 valid data records remained for evaluation. %This chapter describes the evaluation of the study data, including the validation of the stimulus and the reaction evaluation to the different Memoji moods using traditional and modern SER techniques. Furthermore, user strategies and opinions are evaluated according to the participants’ answers to the questionnaire.
%\subsection{Demographic Data}
%Regarding the demographics, the participants had an average age of 30.48, where the youngest participant was 11 years old and the oldest participant was at the age of 69. The participants were asked to provide their exact age instead of selecting from an age range. Figure 4.1 shows the age of the participants as a histogram. The participants were also asked to choose their gender out of the three options female, male and other. No participant selected the option ’other’ which leaves 22 female and 24 male participants. Therefore, the gender distribution is almost balanced between female and male. The average age of all female participants (30.41) compared to the average age of all male participants (30.54) is almost balanced as well.
The participants, 22 female and 24 male, had an average age of 30.5, where the youngest participant was 11 years old and the oldest participant was at the age of 69. The average age of all female participants (30.4) was nearly equal to the average age of all male participants (30.5).

\begin{figure}[h]
\begin{minipage}[t]{0.45\textwidth}
\centering
  \includegraphics[width=\columnwidth]{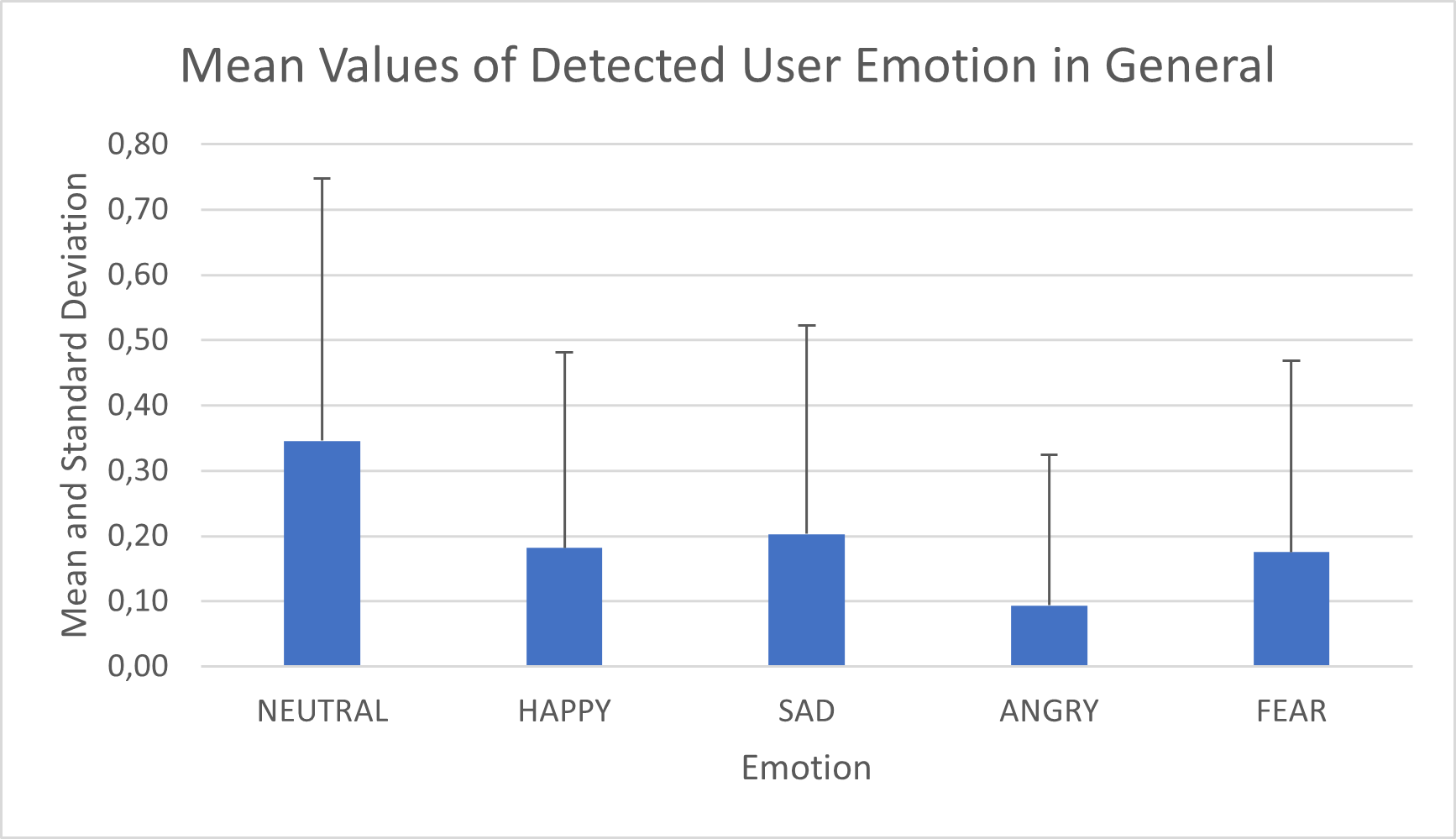}
  \caption{Emotion in the participants' voice (analysed with Vokaturi) averaged over avatar's mood.}
  \label{fig:emotion__in_general}
 %\vspace{-1em}
  \end{minipage}
  \hspace{1em}
\begin{minipage}[t]{0.45\textwidth}
\centering
  \includegraphics[width=\columnwidth]{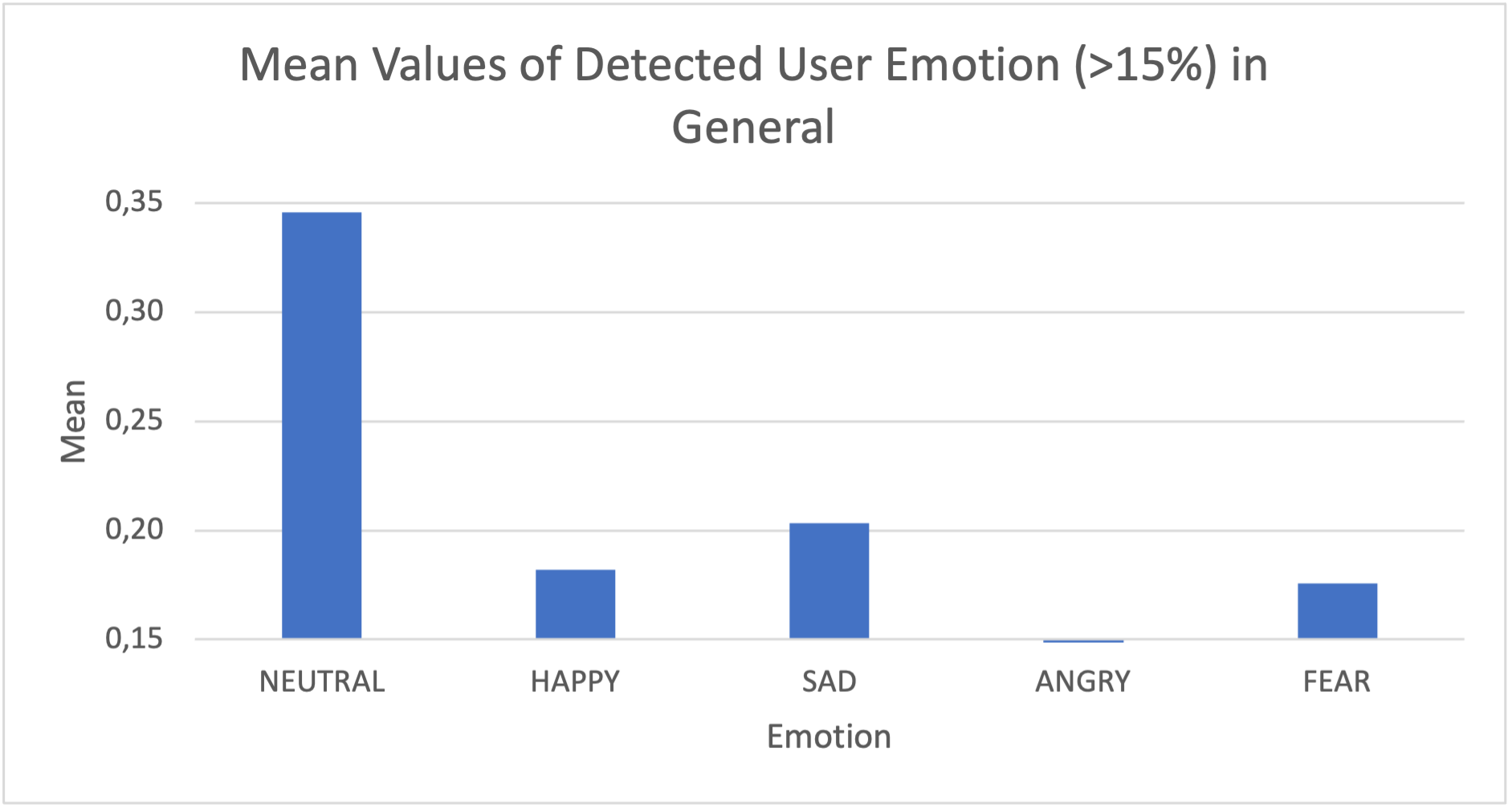}
  \caption{Vokaturi reports at least small values for all emotions, which we consider as noise. The figure above takes only values above 0.15 into account. }
  \label{fig:emotion__in_general_15_percent}
  %\vspace{-1em}
  \end{minipage}
\end{figure}

The evaluation of the users' perception of the stimulus showed that the avatar's mood was perceived as intended. Figure ~\ref{fig:emotion__in_general} shows the users' average emotion as reported by Vokaturi. The high standard deviations indicate a high dispersion of individuals' emotion.
Vokaturi reports five values for five basic emotions and typically four values are small and only one value is high. If we eliminate values below 15\% as noise we get Figure ~\ref{fig:emotion__in_general_15_percent}, which shows that most answers were given with neutral emotion. 

\begin{figure}[h]
\begin{minipage}[t]{0.45\textwidth}
\centering
  \includegraphics[width=\columnwidth]{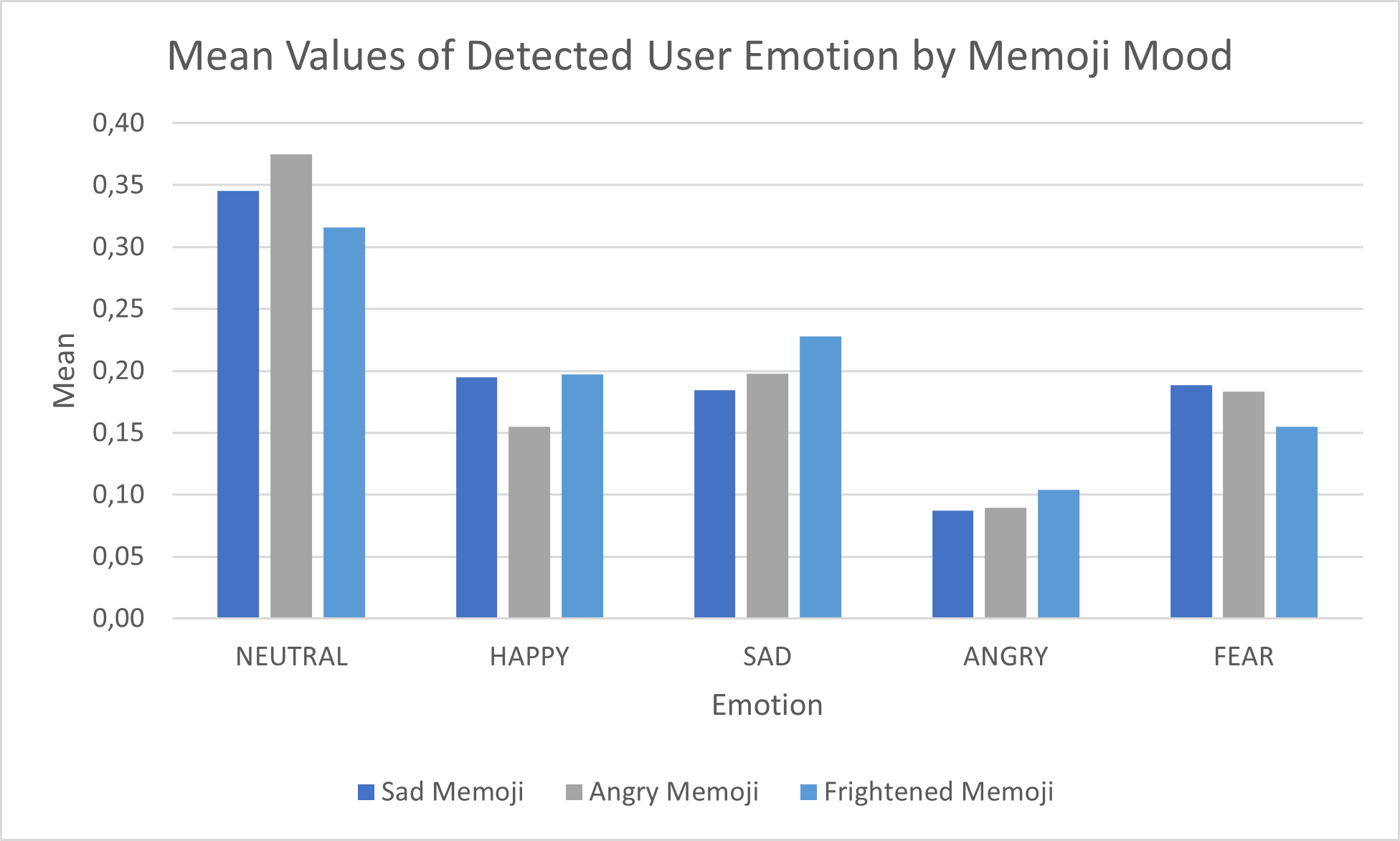}
  \caption{Emotion in the participants voice (analysed with Vokaturi) for the three avatar's mood.}
  \label{fig:emotion_by_avatar_mood}
  %\vspace{-1em}
  \end{minipage}
  \hspace{1em}
\begin{minipage}[t]{0.45\textwidth}
\centering
  \includegraphics[width=\columnwidth]{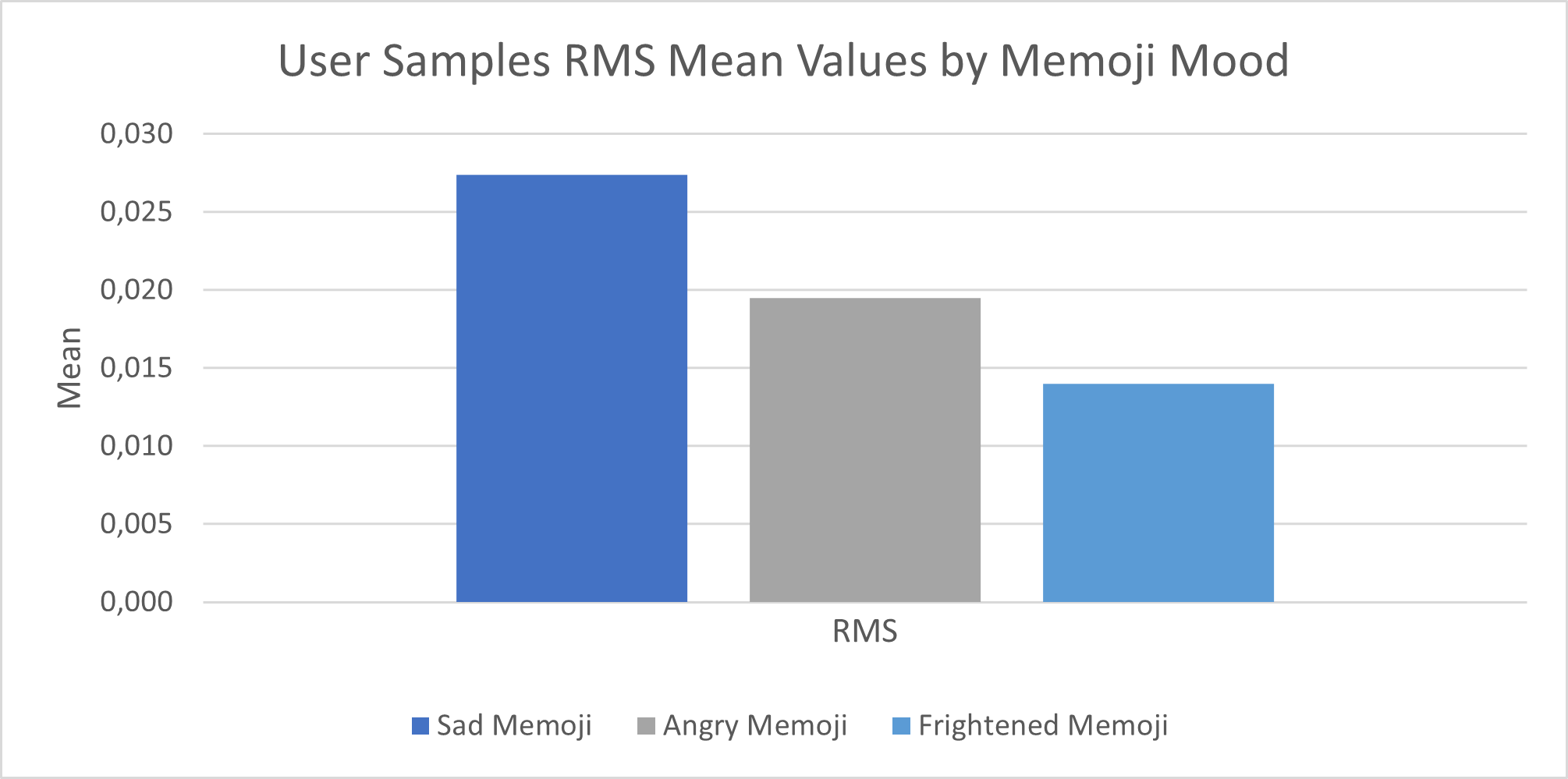}
  \caption{Mean RMS values for voice samples for the different avatar moods.}\label{fig:Mean_RMS_by_mood}
  %\vspace{-1em}
  \end{minipage}
\end{figure}

\begin{figure}
%\begin{minipage}[t]{0.45\textwidth}
\centering
  \includegraphics[width=0.5\columnwidth]{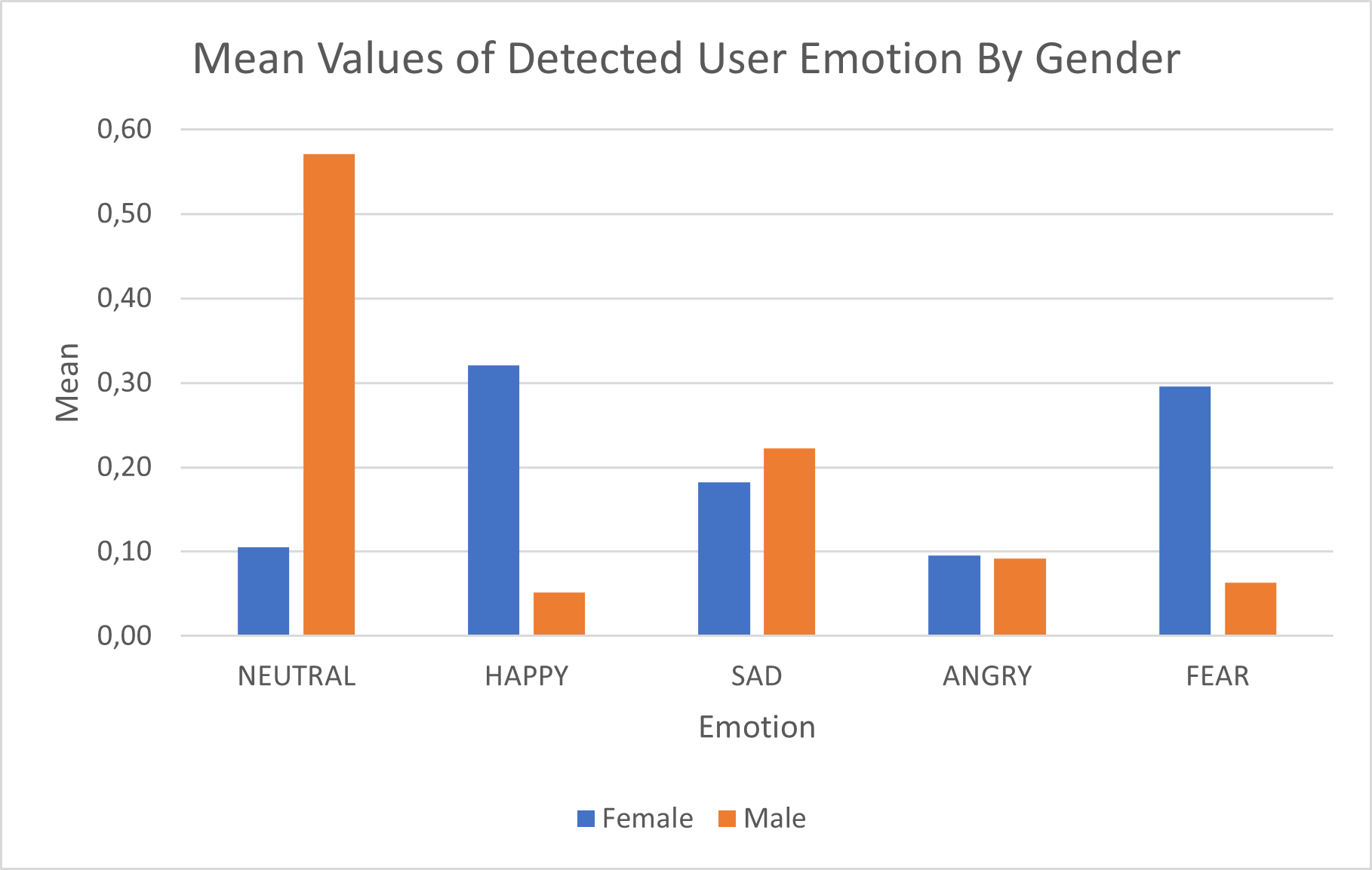}
  \caption{Emotion in the participants voice over gender (analysed with Vokaturi) There is a clear gender difference. }\label{fig:emotion_by_gender}
  %\vspace{-1em}
%  \end{minipage}
\end{figure}

Figure ~\ref{fig:emotion_by_avatar_mood} shows the users' average emotion for the three negative avatar moods. There is not much difference in the emotional response to the different avatar's mood. However, we found a difference in the RMS (root mean square) (see Figure ~\ref{fig:Mean_RMS_by_mood}).
Figure \ref{fig:emotion_by_gender} shows the users' emotion by gender averaged over the three different avatar moods. It shows that there is a clear difference by gender. We did not calculate the significance of this result as it is not clear how reliable Vokaturi's results are. Vokaturi's documentation states "The accuracy on the five built-in emotions is 76.1\%"\footnote{https://developers.vokaturi.com/q-a}.

\section{Discussion and Future Work}

Emotion in speech signals is independent from language and culture \cite{Pell2009} and there was no reason to assume a difference for gender. The fact that we got a gender difference needs an explanation. Although it is unlikely, it could be by chance. 
 
Another explanation could be a gender bias in the emotion detector. Vokaturi's training databases\footnote{https://developers.vokaturi.com/algorithms/annotated-databases} are the Berlin Database of Emotional Speech or Emo-DB\footnote{http://www.expressive-speech.net/emodb/} \cite{Burkhardt2006}, which contains five female and five male speakers, and SAVEE\footnote{http://kahlan.eps.surrey.ac.uk/savee/}, which has voice samples of four males. This raises the question whether the training databases for emotion detection have to be gender-balanced. Vogt et al. showed that gender differentiation improves emotion recognition~\cite{Vogt2006}.

A further possibility is that there is a gender difference in the reaction to emotion. The consequence would be, that a female voice assistant should have a different reaction to emotion than a male voice assistant. It may be the case that male users need a different treatment to cheer them up than female users. Although there are hints that this could be the case \cite{Soobin2022}, we are not very comfortable with this idea, as it means that emotion-aware voice assistants also need gender-awareness and this would manifest gender differences in technology. The general question is whether there is a difference in conversations from man to man, women to woman, or man to woman, and if there are differences, whether we want to implement this in voice assistants.

There are critical voices on gender issues for voice assistants in a UNESCO report\footnote{https://unesdoc.unesco.org/ark:/48223/pf0000367416} and in the media\footnote{https://www.nytimes.com/2019/05/22/world/siri-alexa-ai-gender-bias.html}. There is also an initiative for a genderless voice to end gender bias in AI\footnote{https://www.genderlessvoice.com/}.
A follow-up study could offer a male and a female stimulus. Alternatively, it would be possible to offer a gender-neutral human voice or even the voice of a comic character. 
%The question, which option to prefer, depends whether humanity want to have voice assistants with or without gender.
This raises a more general questions: Should research on future voice interfaces investigate all interfaces which are \emph{possible}, only those for which there is a \emph{market}, or only interfaces which are in accordance with current \emph{morals}?

\bibliographystyle{ACM-Reference-Format}
\bibliography{workshop_draft}

\end{document}